\documentclass[aps,prb,reprint,groupedaddress,superscriptaddress,showpacs,floatfix]{revtex4-1}
\usepackage{braket}
\usepackage{graphicx}
\usepackage{dcolumn}
\usepackage{bm}
\usepackage{epsfig}
\usepackage{amsmath}
\usepackage[usenames,dvipsnames]{color}
\usepackage{booktabs}
\usepackage{hyperref}
\usepackage{color}
\usepackage{nccmath}
\usepackage{lipsum}

\usepackage{mathrsfs}
\usepackage{appendix}
\usepackage{bbm}

\def \vn {{\bf n}}

\def \vS {{\bf S}}
\def \vR {{\bf R}}

\def \mb {\mu_{\rm B}}

\def \vB {{\bf B}}

\def \va {{\bf a}}
\def \vb {{\bf b}}
\def \vc {{\bf c}}

\def \tig {{\rm Tb$_2$Ir$_3$Ga$_9$}\,\,}

\def \vn {{\bf n}}

\begin{document}

\title{Steplike metamagnetic transitions in a honeycomb lattice antiferromagnet Tb$_2$Ir$_3$Ga$_9$}

\author{Mojammel A. Khan}

\email{mkhan19@anl.gov, mojammelkhan1987@gmail.com}
\affiliation{Materials Science Division, Argonne National Laboratory, 9700 South Cass Avenue, Argonne, Illinois 60439, USA}

\author{Qiang Zhang}
 \affiliation{Neutron Scattering Division, Oak Ridge National Laboratory, Oak Ridge, Tennessee 37831, USA}

\author{Jin-Ke Bao}
\affiliation{Materials Science Division, Argonne National Laboratory, 9700 South Cass Avenue, Argonne, Illinois 60439, USA}

\author{Randy S. Fishman}
\affiliation{Materials Science and Technology Division, Oak Ridge National Laboratory, Oak Ridge, Tennessee 37831, USA}


\author{A. S. Botana}
\affiliation{Department of Physics, Arizona State University, Tempe Arizona 85281, USA}

\author{ Y. Choi}
\affiliation{Advanced Photon Source, Argonne National Laboratory, Argonne Illinois 60439, USA}
\author{G. Fabbris}
\affiliation{Advanced Photon Source, Argonne National Laboratory, Argonne Illinois 60439, USA}
\author{D. Haskel}
\affiliation{Advanced Photon Source, Argonne National Laboratory, Argonne Illinois 60439, USA}

\author{John Singleton}
\affiliation{National High Magnetic Field Laboratory, Los Alamos National Laboratory, MS-E536, Los Alamos, New Mexico 87545, USA}

\author{J. F. Mitchell}
\affiliation{Materials Science Division, Argonne National Laboratory, 9700 South Cass Avenue, Argonne, Illinois 60439, USA}

\date{\today}
\begin{abstract} 

Single crystals of a honeycomb lattice antiferromagnet, Tb$_2$Ir$_3$Ga$_9$ were synthesized, and the physical properties have been studied. From magnetometry, a long-range antiferromagnetic ordering at $\approx$12.5 K with highly anisotropic magnetic behavior was found. Neutron powder diffraction confirms that the Tb spins lie along the $\va $-axis, parallel to the shortest Tb-Tb contact. Two field-induced spin-flip transitions are observed when the field is applied parallel to this axis, separated by a plateau corresponding roughly to M$\approx$M$_{\rm{s}}$/2.  Transport measurements show the resistivity to be metallic with a discontinuity at the onset of N\'eel order. Heat capacity shows a $\lambda$-like transition confirming the bulk nature of the magnetism. We propose a phenomenological spin-Hamiltonian that describes the magnetization plateau as a result of strong Ising character arising from a quasidoublet ground state of the Tb$^{3+}$ ion in a site of \textit{C$_s$} symmetry and expressing a significant bond dependent anisotropy.

\end{abstract}

\maketitle

\section{\label{sec:level1}Introduction}

Materials containing honeycomb lattices decorated by metals with strong spin-orbit coupling are potential hosts for Kitaev quantum spin liquids (QSL)~\citep{kitaev2006anyons,PhysRevLett.105.027204}. The essential ingredient for understanding magnetism in these systems is a dominant bond-directional anisotropic exchange that leads to magnetic frustration~\citep{bond.directional.kitaev,bond.directional.kitaev2}. Recent attention to $4d$ and $5d$ transition metal-based candidates such as $\alpha$-RuCl$_3$~\citep{RuCl3} and A$_2$IrO$_3$ (A=Li, Na)~\citep{PhysRevLett.105.027204,Iridate} underscores the interest in this honeycomb structural motif, with the bond-directional anisotropy a consequence of the spin-orbit coupling derived $j_{\rm{eff}}$=1/2 ground state assigned to the transition metals. Decoration of the honeycomb lattice with rare-earth ions offers an alternative to $4d$- and $5d$-based materials, with YbCl$_3$ (isostructural with $\alpha$-RuCl$_3$) suggested as a potential Kitaev QSL candidate~\citep{xing2019rare}.  Addionally, a recent theoretical treatment of spin-orbital entanglement in rare-earth honeycomb magnets by Luo and Chen~\citep{rare.kitaev}, and others~\citep{kitaev.afm,clark2019two} highlights the need to explore such systems. 

A nearly ideal honeycomb lattice of rare earth ions is found in a family of compounds with general formula, R$_2$T$_3$X$_9$ ( \textit{R} is a rare-earth element, \textit{T} is a transition metal element and \textit{X} is a $p$-block element)~\citep{Y2Co3Ga9,EuIrGa_structure}. This family, typified by the Y$_2$Co$_3$Ga$_9$ structure~\citep{Y2Co3Ga9} occupies a large composition space and hosts a rich variety of electronic properties ranging from complex magnetically ordered states (Dy based compounds), mixed valence (Yb,Ce based compounds) and Kondo lattice behavior (Yb based compounds) ~\citep{DyCoAl,Ce2Pd9Sb3,Yb2Co3T9,Yb2RhIR3AlGa9,R2Co3Al9,Yb2.Konod.Lattice,U2Co3Al9,niermann2004preparation}.

 The layered structure of R$_2$T$_3$X$_9$, along the [001] direction can be viewed as a stacking of two alternating types of layers~\citep{EuIrGa_structure} (Fig.~\ref{figure 1}). The \textit{R} and \textit{X} atoms form a planar layer with the composition R$_4$X$_6$, which contains a two-dimensional honeycomb-like arrangement of \textit{R} atoms. The other layer is strongly puckered with \textit{T} and \textit{X} atoms forming a hexagonal arrangement, with \textit{T}:Ga ratio of
1 : 2 (T$_6$X$_{12}$). Overall stacking is such that the resulting structure is orthorhombic with space group \textit{Cmcm} no. 63. The physical properties of these compounds depend on identity of \textit{X} and \textit{T} atoms that surround the \textit{R} atom. Depending on the ligand \textit{X} atoms, systems with the same R atom can be magnetic in some cases and non-magnetic in others~\citep{DyCoAl}. It is evident that the Al-containing compounds Yb$_2$T$_3$Al$_9$ are magnetic, whereas their Ga-containing analogues are paramagnets down to the lowest temperatures measured~\citep{DyCoAl,Yb2Co3T9,Yb2RhIR3AlGa9}. Thus, the interactions between \textit{R} atoms via the RKKY mechanism coupled with the large magnetocrystalline anisotropy due to crystal electric field (CEF) acting on \textit{R} atom could lead to complex magnetic states in this family of compounds. Indeed, multiple commensurate and incommensurate phases as well as field induced metamagnetic states have been reported~\citep{DyCoAl}. As such, this family of compounds presents an opportunity to understand the interplay between various hybridization strengths of different \textit{d}, and \textit{f} electrons and ligand atoms in a layered structure where the magnetic \textit{R} atoms form a slightly distorted honeycomb lattice.

Interestingly, Tb containing compounds, such as Tb$_2$T$_3$X$_9$, have not been explored and in this article we present our study on single crystalline \tig. Magnetization measurements reveal a long-range antiferromangetic ordering at 12.5 K. The magnetic susceptibility is highly anisotropic, with strong preference for in plane magnetization. Two field-induced metamagnetic transitions separated by a plateau of M$\approx$M$_{\rm{s}}$/2, where M$_{\rm{s}}$ is the saturation moment, are observed when the field is applied along the crystallographic $\va $-axis, which is also the magnetic easy axis. A finite ferromagnetic-like response was observed along the $\vb $-axis, although neither $\vb $ nor $\vc $-axes show field-induced metamagnetic transitions. Neutron powder diffraction supports the observed magnetic behaviors by revealing the zero field ground state to be a collinear two-sublattice antiferromagnet (AFM) with spins along the $\va $ axis. Based on observed properties, we present a phenomenological spin model to describe the magnetic behavior  for \tig.  Notably, incorporation of a bond-directional anisotropy term with significant weight, is essential for agreement between theory and experiment, making contact with the phenomenology of transition-metal based honeycomb magnets.

This article is arranged as follows. In section II, experimental methods and processes are described. In section III, results of magnetization, transport, specific heat, X-ray magnetic circular dichroism (XMCD), neutron powder diffraction (NPD) measurements, and the electronic structure calculations are presented. In section IV, a phenomenological theory of the magnetization in Tb$_2$Ir$_3$Ga$_9$ is presented. Finally, in section V we conclude with discussion of the ground state properties of Tb$_2$Ir$_3$Ga$_9$.

\section{\label{sec:level2}Experimental description}

Single crystals of Tb$_2$Ir$_3$Ga$_9$ were synthesized using a Ga-flux method. The starting materials, Tb pieces (Alfa Aesar, 99.9$\%$), Ir powder (Alfa Aesar, 99.99$\%$), and Ga pellet (Alfa Aesar, 99.9999$\%$) in the molar ratio of 1:2:20, were placed in an alumina crucible and sealed under vacuum in a fused silica tube. The ampoule was heated to 1170$^{\circ}$ C and held for 12 hours, then cooled to 500$^{\circ}$C at a rate of 5$^{\circ}$C per hour. The excess flux was removed using a centrifuge. The resultant crystals are in the form of hexagonal platelets with an average size of few millimeters (mm) on an edge, as shown in Fig.~\ref{figure 1}c.

One large piece of \tig single-crystal was cut into smaller pieces with appropriate dimensions ($\approx0.1\times0.1\times0.1 {\rm\, mm}^3$) for single crystal x-ray diffraction measurements. A tiny piece of crystal was mounted on a glass fiber and measured on a STOE IPDS 2T. Data collection, integration and absorption correction were done by the x-area software package~\citep{stoe2005x}. The structure of \tig was solved and further refined based on the full matrix least-squares  using the SHELXTL program package~\citep{sheldrick2015crystal}. An empirical absorption correction was applied to the measured data. The refinement results including the lattice parameters and atomic positions are consistent with the previous reports on a polycrystalline sample~\citep{grin1989phases}. Several pieces of crystals were pulverized and powder XRD was performed on a PANAlytial X'Pert Pro diffractometer. Magnetization measurements were carried out in a Quantum Design SQUID magnetometer using the DC magnetization method. For susceptibility, both zero-field-cooled (ZFC) and field-cooled (FC) data were measured. Isothermal magnetization data were measured  by first cooling the sample in zero field to 1.8 K and then applying field to $\pm$ 7 T. 

Transport measurements were performed in a Quantum Design Physical Property Measurement Systems (PPMS).  A piece of crystal was polished to a rectangular shape with dimensions 1$\times$0.6$\times$0.1 mm$^3$ and then gold wire (25 micron) contacts were placed using Epotek H20E Epoxy. A four-probe contact method was used for the AC resistivity measurement with an excitation current of 3-5 mA at a frequency of 57.9 Hz. Heat capacity measurement was performed in a Quantum Design Dynacool PPMS. A time relaxation method was employed and the data were measured on heating from 1.8 K to 200 K under zero applied field. 

The angular dependence of the metamagnetic transition was measured in a 65 T magnet at the Pulsed-Field Facility, National High Magnetic Field Laboratory, Los Alamos~\citep{singleton2004national}. The metamagnetic transitions were measured in two rectangular samples using the Proximity-Detector Oscillator (PDO) method. Samples were rotated from field parallel to crystallographic $\va$-axis to $\vc \& \vb$ axes. A more in depth discussion about the PDO technique can be found in Refs.~\citep{PDO.1,PDO.2}.

Neutron powder diffraction (NPD) measurements were 
performed on the time-of-flight powder diffractometer,
POWGEN, located at the Spallation Neutron Source at Oak
Ridge National Laboratory. The data were collected with neutrons
of central wavelengths 1.5 and 2.665 \AA, covering the $Q$ spacing
range 0.48$-$12.95 and 0.3$-$5.87 \AA$^{-1}$, respectively. Several high-quality crystals were pulverized to obtain around 0.55 g powder that was loaded in a special annular vanadium container to reduce the 
absorption effect from Tb and Ir and gain more diffraction intensity.
A Powgen Automatic Changer (PAC) was used to cover the temperature region of 
10$-$300 K. We collected the data at 10, 50 and 300 K. All of the neutron diffraction data were analyzed
using the Rietveld refinement program suite FULLPROF~\citep{Carva1993}.

XMCD data were collected at beamline 4-ID-D of the Advanced Photon Source (APS). Two single crystals with surface normals along the crystallographic $\va $- and $\vb $- directions were polished to $\approx$20 microns thickness for transmission experiments at Tb L$_3$ and Ir L$_{2,3}$ absorption edges. Crystals were mounted in a variable temperature insert of a cryogenic superconducting magnet and cooled to 1.5 K in helium vapor. Magnetic field was applied parallel to $\va $- or $\vb $-axis. XMCD data were collected in helicity switching mode (fixed magnetic field direction) whereby the helicity of circularly polarized x-rays produced with phase retarding optics is modulated at 13.1 Hz, and the corresponding modulation in x-ray absorption coefficient is detected with a phase lock-in amplifier~\citep{haskel2007instrument}. XMCD measurements were done with applied magnetic field both along and opposite the wave vector of the incident x-ray beam to check for experimental artifacts of non-magnetic origin.

\section{\label{sec:level3}Results \& discussion}

\subsection*{Crystal Structure}

\begin{figure}[ht]
\centering
\includegraphics[width=0.5\textwidth]{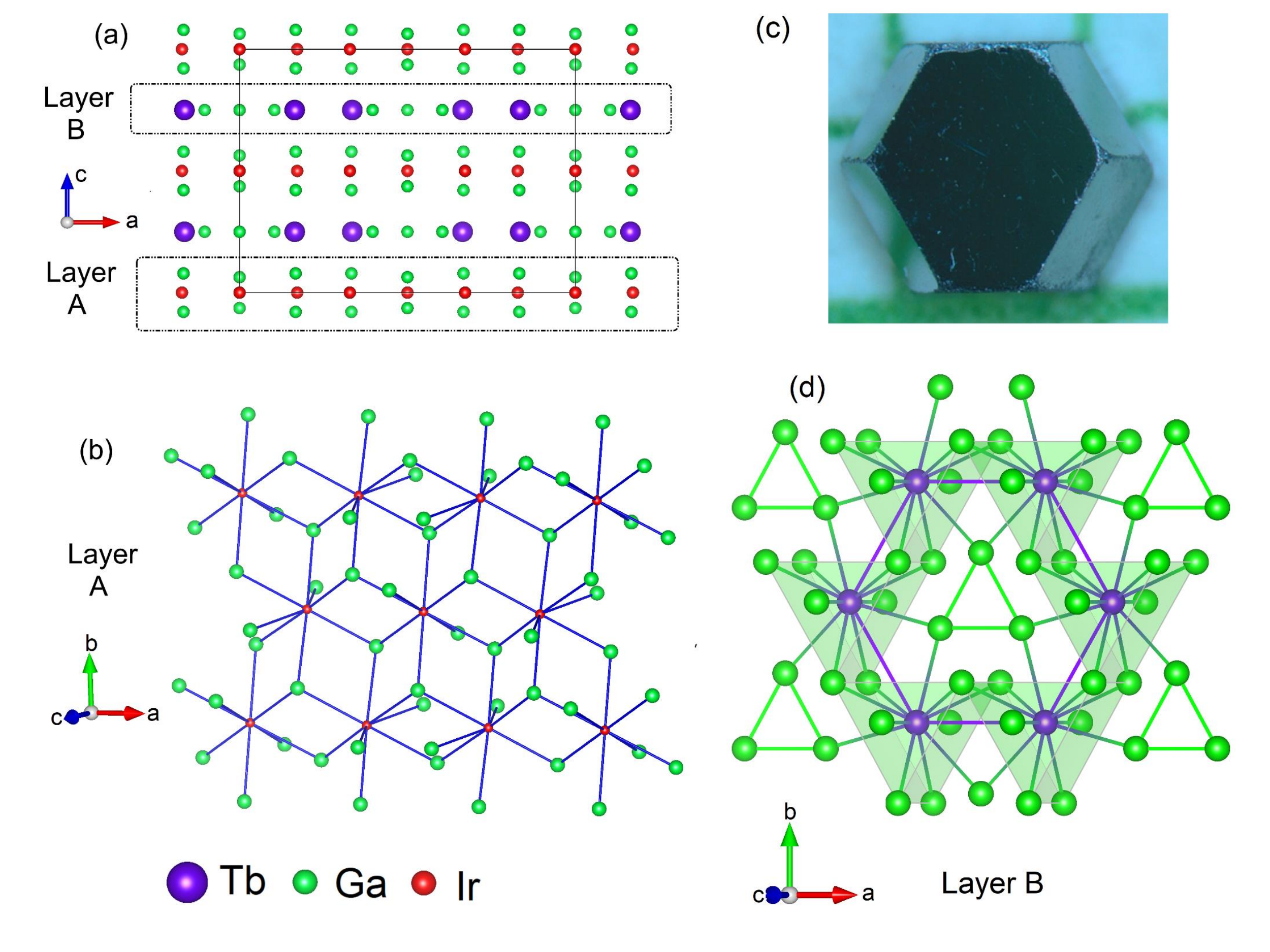}
\caption{Crystal structure of Tb$_2$Ir$_3$Ga$_9$.(a) Schematic of crystal structure projected along \textit{ac} plane. Two different layers are identified that make up the structure in stacking sequence (AB)$_2$. (b) Layer A projected as described in the image. The Ir-Ga coordination is Ir$_6$Ga$_{12}$.(c) As grown crystal of Tb$_2$Ir$_3$Ga$_9$ on a 1 mm grid. (d) Layer B projected as indicated in the figure. The honeycomb arrangement of Tb atoms is evident.}\label{figure 1}
\end{figure}

The crystal structure of Tb$_2$Ir$_3$Ga$_9$ is orthorhombic, crystallizing in space group \textit{Cmcm} (no. 63), and is isotypic with  Y$_2$Co$_3$Ga$_9$~\citep{Y2Co3Ga9}. As shown in Fig.~\ref{figure 1}, the structure can be viewed as an alternation of two layers along the [0 0 1] direction. Layer A is strongly puckered with a hexagonal arrangement and consists of Ir and Ga in a 1:2 ratio. Layer B lies in a mirror plane with the Tb atoms arranged in a slightly distorted honeycomb structure. The Tb-Ga coordination in a unit cell is Tb$_4$Ga$_6$. Together these layers form the crystal in a stacking sequence of (AB)$_2$. The ratio of lattice parameters a/b = 1.725, close to $\sqrt{3}$, reflecting a small deviation from hexagonal symmetry. This is similar to other R$_2$T$_3$Ga$_9$ compounds with Y$_2$Co$_3$Ga$_9$ structure type~\citep{EuIrGa_structure, R2T3Al9_RhIrPd,DyCoAl}.

\begin{table*}[ht]\caption{Crystal data and structure refinement for Tb$_2$Ir$_3$Ga$_9$ at 293 K.}
\centering 
\small\addtolength{\tabcolsep}{-4pt}
\begin{tabular}{cc}
\hline
Empirical Formula  & Ga$_9$Ir$_3$Tb$_2$  \\

Formula Weight & 1521.92 g/mol\\
Wavelength & 0.71073  \AA \\

Crystal System and Space Group& Orthorhombic, $Cmcm$   \\

Unit Cell Dimensions& a = 12.9860(5) \AA \\
& b = 7.5325(9) \AA \\
& c = 9.4349(9) \AA \\
& $\alpha$ = $\beta$ = $\gamma$ = 90$^\circ$ \\

Volume & 922.89(2) \AA$^3$\\
Density (calculated) & 10.953 mg/m$^3$\\
Linear Absorption Coefficient& 83.867 mm$^{-1}$\\

F(000) & 2560 electrons\\
Crystal Size & 0.11$\times$0.1$\times$0.95 mm$^3$\\
$\theta$ Range &	3.126 to 31.843.\\
Index Range & -19$<$=h$<$=18, -11$<$=k$<$=11, -13$<$=l$<$=13\\
No. of Reflections & 5326\\
Independent Reflections & 857[R$_{int}$  = 0.1140]\\
Absorption Correction	& empirical\\
Max. and min. Transmission &	0.0786 and 0.0176\\
Refinement Method & Full-matrix least-squares on F$^2$\\
Data / Restraints / Parameters &	857 / 0 / 42\\
Goodness of Fit& 1.151\\
R$_{final}$ indices & 	R$_1$ = 0.0385, wR$_2$ = 0.1006\\
R indices (all data)& 	R$_1$ = 0.0431, wR$_2$ = 0.1038\\
Extinction Coefficient&	0.00066(7)\\
Largest diff. Peak and Hole &	4.766 and -4.565 e.\AA$^{-3}$\\

\hline
 
\end{tabular} 
\label{tab: single crystal xrd}
\end{table*}

The results of the single-crystal x-ray diffraction are given in Table~\ref{tab: single crystal xrd}. A crystal with a shape close to a cube (0.11$\times$ 0.1$\times$ 0.095 mm$^3$) gives the lowest R$_{int}$ (11\%) and reasonable thermal displacements during the refinement. The large residual electron peaks and holes close to the heavy elements Ir and Tb atoms are likely due to an inadequate absorption correction. Detailed descriptions of the atomic coordinates, refinement parameters and bond lengths of the atoms are given in the supplemental materials (SM)~\citep{supp}.

\subsection*{Magnetism \lowercase{and} Transport}

\begin{figure}[]
\centering
\includegraphics[width=0.4\textwidth]{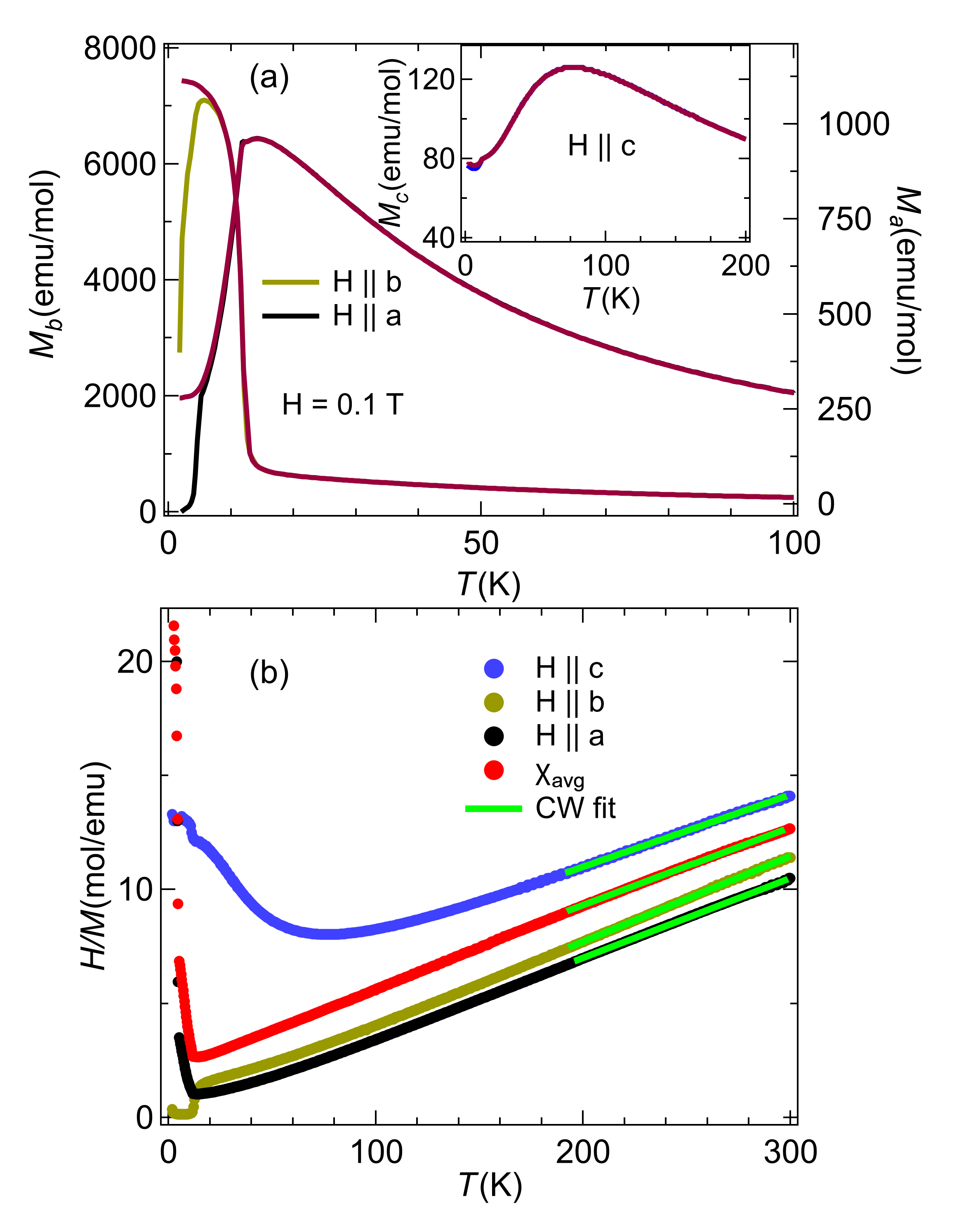}
\caption{DC Magnetization of Tb$_2$Ir$_3$Ga$_9$.(a) \textit{M(T)} with field applied along the $\va- $ and $\vb- $ axes (Inset: \textit{M(T)} along $\vc $-axis). The maroon solid line represents the FC data. (b) ZFC inverse susceptibility in an applied field of 0.1 T for three distinct crystallographic axes and the polycrystalline average as described in text. Green solid line represents a fit to the Curie-Weiss form as described in the text.}\label{figure 2 MT}
\end{figure}

\begin{figure}[]
\centering
\includegraphics[width=0.4\textwidth]{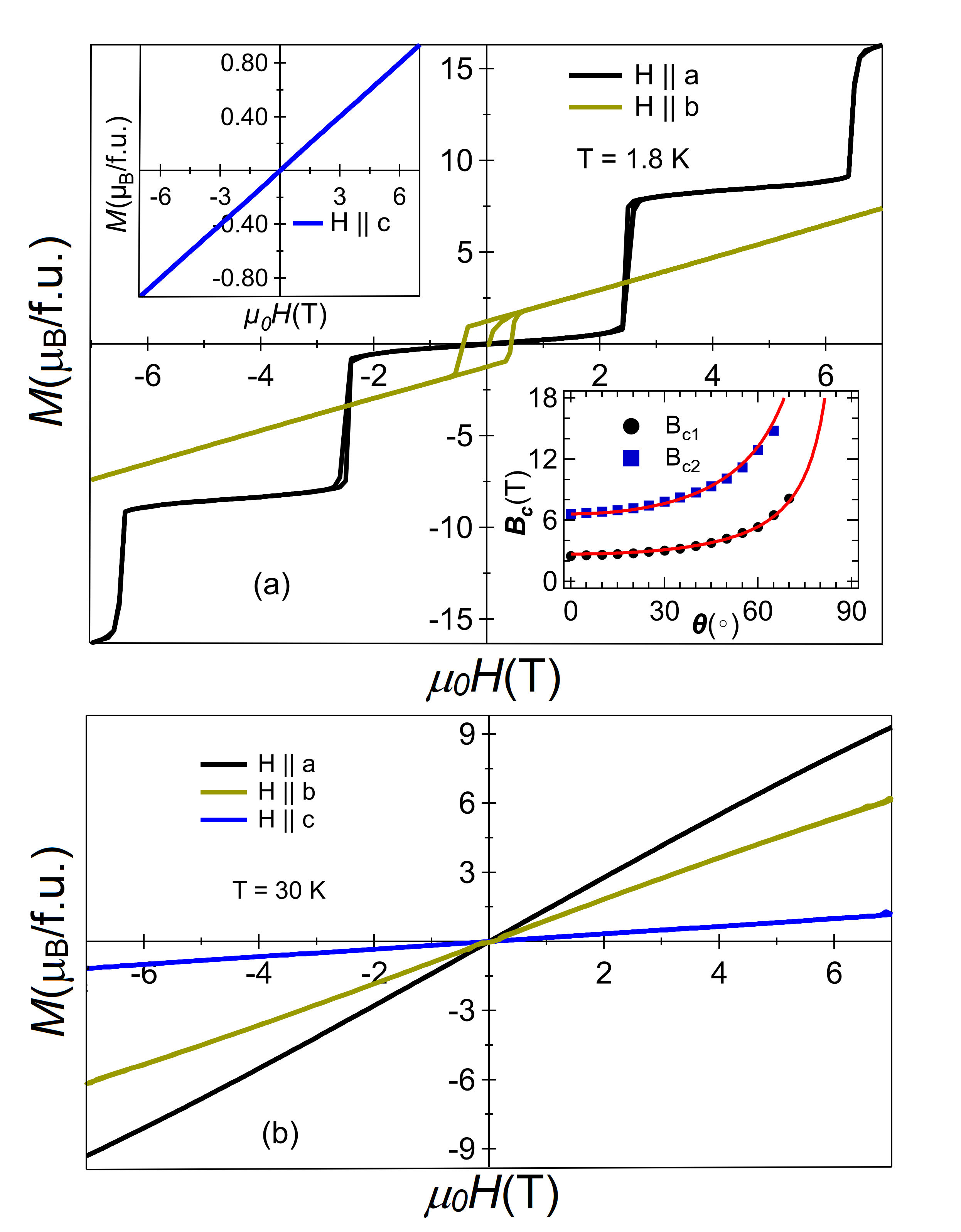}
\caption{Magnetization at 1.8 K and 30 K.(a) Magnetization \textit{M(H)} with field applied along three crystallographic axes at 1.8 K. The \textit{M(H)} along $\va $ axis shows ultra-sharp metamagnetic transitions with each jump corresponding to a value of M$_{\rm s}$/2. Upper left inset: \textit{M(H)} along $\vc $-axis. Lower right inset: Angle dependence of critical field, $B_{\rm c}$, of metamagnetic transitions. The angle $\theta$ is measured from $\va$ toward $\vb$ axis. Red solid lines are fit to the form, $B_{\rm c}/cos\theta$. (b) \textit{M(H)} at 30 K.}\label{figure 3 MH}
\end{figure}

The temperature-and magnetic-field dependent magnetization measurements, \textit{M(T,H)}, are shown in Figs.~\ref{figure 2 MT} \&~\ref{figure 3 MH}, respectively. Magnetism in Tb$_2$Ir$_3$Ga$_9$ is highly anisotropic, as is evident from the DC magnetic susceptibility shown in Fig.~\ref{figure 2 MT}a. The susceptibility along the $\va $-axis shows antiferromagnetic ordering at 12.5 K with ZFC and FC curves showing no irreversibility up to 3 K. In contrast, the susceptibility for field parallel to the $\vb $-axis shows ferromagnetism with a discontinuity around 13 K. The FC curve shows large irreversibility below the transition.

The susceptibility along the $\vc $-axis (Inset of Fig.~\ref{figure 2 MT}a) is characterized by a broad maximum around 65 K followed by two discontinuities at 12.5 K and 2.5 K. The broad maximum is likely attributable to higher-lying CEF states being populated by thermal excitation. While the 12.5 K transition probably marks long-range order, we cannot rule out the possibility that this feature arises from small misalignment of the crystal with respect to the field. Some Ce and Yb based compounds with similar crystal structure show broad peak in temperature ranges 150-250 K, and those are well understood by mixed valence states along with CEF effects~\citep{Ce2T3X9.transport,CEF.Yb2Ir3Ga9,Yb2RhIR3AlGa9}. The weak feature at 2.5 K is currently not understood. The ZFC and FC data only show irreversibility at the onset of the transition. For $T<T_{\rm N}$, susceptibility along $\vb $ is largest followed by $\va $ and $\vc $, while for $T>T_{\rm N}$, the order is $\chi_a>\chi_b>\chi_c$. 

Figure~\ref{figure 2 MT}b shows inverse susceptibility  along all three axes and the average susceptibility, $\chi_{\rm avg} = \frac{(\chi_a +3 \chi_c)}{3}$, with a fit to the Curie-Weiss form, $1/\chi$ = $\frac{T-\theta_{\rm W}}{C}$, shown as a solid line. The Weiss temperature, $\theta_{\rm W}$, and effective moment, $\mu_{\rm eff}$, estimated from the Curie constant,\textit{C}, provide insights on the interactions between magnetic atoms. Different values of Weiss temperature reflect the strong anisotropy in magnetization and are similar to other compounds of the R$_2$T$_3$X$_9$ family~\citep{DyCoAl}. The value of $\theta_{\rm W}$ is largest for the field along $\vc $ with a value of -148$\pm$2 K followed by  -5.8$\pm$0.3 K along $\vb $ and -0.5$\pm$0.1 K along $\va $ axis. In all three directions, $\theta_{\rm w}<0$, indicating antiferromagnetic interactions of varying strengths. The average effective moment estimated from the fit to $\chi_{\rm avg}$, is 10.3 $\mu_{\rm B}$/Tb, close to that expected effective moment for a free Tb$^{3+}$ ion $\approx$9.7 $\mu_{\rm B}$, while the average Weiss temperature is -70 K.



The anisotropic magnetism in Tb$_2$Ir$_3$Ga$_9$ is also evident in \textit{M(H)} data, shown in Fig.~\ref{figure 3 MH}a for 1.8 K, and in Fig.~\ref{figure 3 MH}b for 30 K. When the field is applied along the $\va $-axis, two field-induced sharp metamagnetic transitions are seen at fields $B_{\rm c1}$ and $B_{\rm c2}$, with each transition corresponding to a jump of M$_{\rm{s}}$/2 $\approx$8 $\mu_{\rm B}$, where M$_{\rm s}$ is the saturated magnetic moment at 1.8 K and 7 T. We note that the expected M$_{\rm s}$ for a free Tb$^{3+}$ ion, $^7$F$_6$, with $g=1.5$ is 9 $\mu_{\rm B}$/Tb, thus the jumps are very close to half of that expected for local Tb$^{3+}$ moment. At 7 T and 1.8 K, the $M$=16.3 $\mu_{\rm B}$/formula unit (f.u.). The angle dependence of the metamagnetic critical field, $B_{\rm c}$, shown in the right side inset of Fig.~\ref{figure 3 MH}(a), implies that only the $\va$-axis projection of field is important. Both transitions move to higher fields with increasing angle. Beyond 75 degrees, the transitions have moved beyond the accessible field range. The solid lines are the fit to the form $B_{\rm c}$/cos$\theta$, demonstrating that both metamagnetic transitions depend only on the component of the field along the $\va $-axis.

Steplike metamagnetic transitions have been found in certain phase-separated perovskite manganites~\citep{metamag.manganite.1,metamag.manganite.2}, and intermetallic compounds such as Nd$_5$Ge$_3$\citep{Nd5Ge3.1,Nd5Ge3.2}, Gd$_5$Ge$_4$~\citep{Gd5Ge4.1} doped CeFe$_2$~\citep{CeFe2}, and LaFe$_{12}$B$_6$~\citep{LaFe12B6}. In these compounds transitions are driven by field-induced response of the phase-separated state, where the applied field favors the ferromagnetic phase over the AFM phase. The metamagnetic transition in these inhomogeneous systems is first-order and is accompanied by a large hysteresis with remanance. In contrast, \tig is a homogeneous AFM system similar to TbNi$_2$Ge$_2$~\citep{TbNi2Ge2}, TbCo$_2$Si$_2$,\citep{DyCo2Si2.1,rare.earth.metamagnetism}, TbCoGa$_5$~\citep{TbCoGa5}, TbCo$_2$Ge$_2$~\citep{TbCo2Si2}, for example. The magnetic behavior follows from the expected Ising character of Tb$^{3+}$ found in the low site symmetry, $C_s$~\citep{XMCD.TbMn03,CEF.TbAl03}. Tb$^{3+}$ is a non-Kramers’ ion, and the ground state manifold $^7F_6$ is split into 2$J$ + 1 = 13 singlets by CEF in this $C_s$ symmetry. The lowest crystal-field level is expected to be a quasidoublet that dictates the low temperature properties of the host compound, including a magnetic moment of 9 $\mu_{\rm B}$~\citep{XMCD.TbMn03,CEF.TbAl03,zvezdinmodern} and the Ising behavior. The observed metamagnetic transitions of \tig are consistent with this scenario with the Ising axis being the crystallographic $\va $-axis. 

While the \textit{M(H)} along $\vc $ axis is typical of an antiferromagnet, the $\vb $ axis magnetization reveals a small hysteresis loop with coercivity of $\approx$ 0.5 T indicating ferromagnetism in accordance with \textit{M(T)} data and a breakdown of the purely Ising approximation. We show in section~\ref{model} below that a Dzyaloshinskii- Moriya (DM) type interaction between Tb$^{3+}$ gives rise to this FM component. The M$_{\rm s}$ at 7 T and 1.8 K along $\vb $-axis is 7.5 $\mu_{\rm B}$/f.u., and the remanant moment is 1.2 $\mu_{\rm B}$/f.u.  

The temperature dependent resistivity, $\rho$, with current applied along three different crystallographic axes is shown in Fig.~\ref{figure 4 resistivity}a. The resistivity is metallic, anisotropic and shows a discontinuity at the magnetic transition. Transport anisotropy, $\rho_{\bf c}/\rho_{\bf a}$ $ \approx$2.5 at 290$\sim$K. The residual resistivity ratio, $\rho_{\rm 290\,K}/\rho_{\rm 2\,K}$ , is about 6 for all axes.


\subsection*{Heat Capacity}

\begin{figure}[]
\centering
\includegraphics[width=0.5\textwidth]{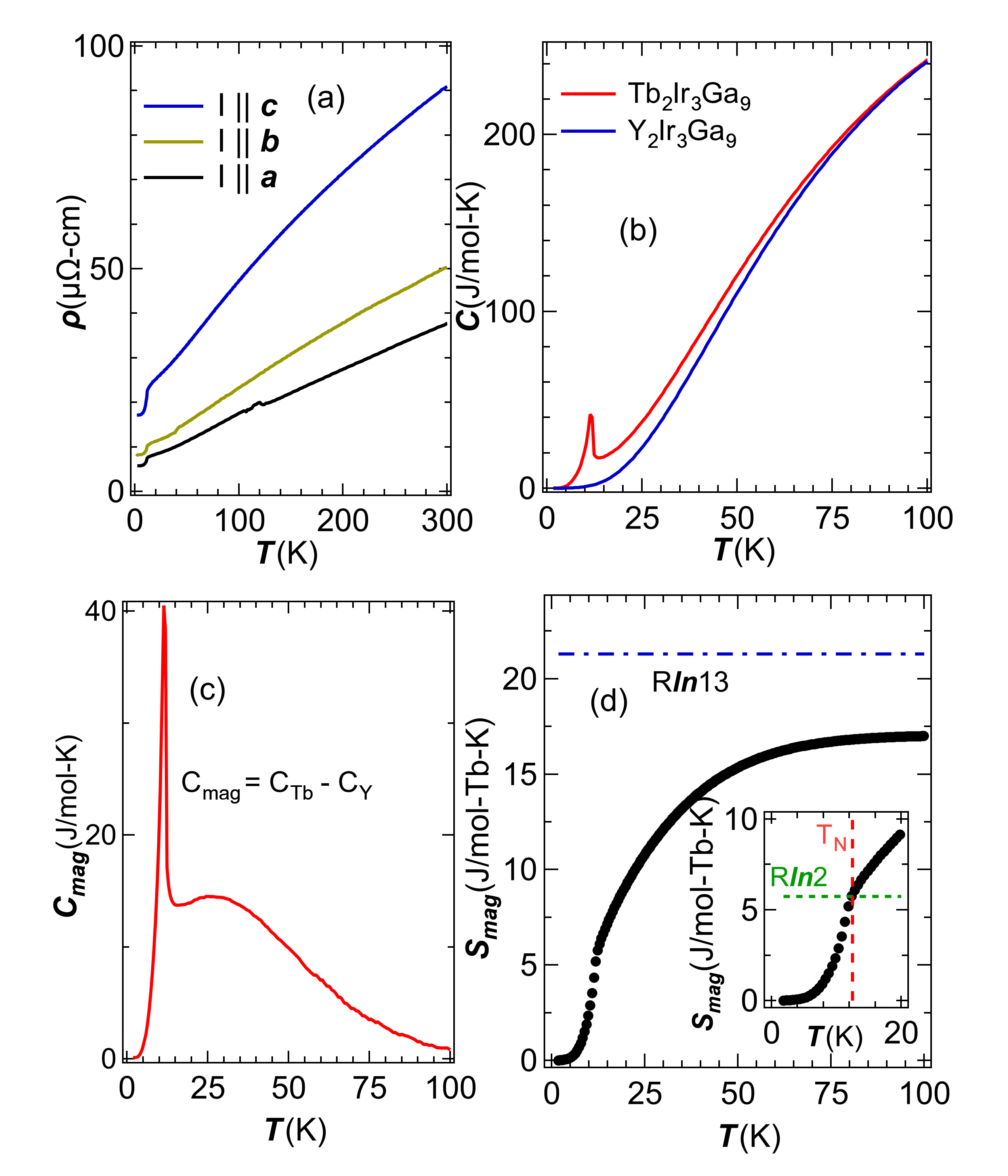}
\caption{(a) Resistivity with current applied along three crystallographic axes. The anisotropy between $\va $ and $\vc $ axis  resistivity at 290 K is close to 2.5. (b) Specific heat, \textit{C}, of Tb$_2$Ir$_3$Ga$_9$ and its non-mangnetic counterpart Y$_2$Ir$_3$Ga$_9$ (blue line) versus temperature, \textit{T}. (c) Magnetic specific heat, \textit{C$_{\rm mag}$}, plotted against temperature.(d) Magnetic entropy, \textit{S$_{\rm mag}$}, per Tb versus temperature. \textit{S$_{\rm mag}$} was calculated from the magnetic specific heat as described in the text. Inset: low temperature region of the entropy.  The horizontal dashed line (olive) is R$ln$2, which coincides with the vertical dashed line (red) marking the T$_{\rm N}$ at the inflection point.}\label{figure 4 resistivity}
\end{figure}

The temperature dependence of the heat capacity, \textit{C$_p$}, of Tb$_2$Ir$_3$Ga$_9$ is shown in Fig.~\ref{figure 4 resistivity}b. The $\lambda$-like anomaly in the heat capacity at 12.5 K denotes the bulk, long-range magnetic ordering in Tb$_2$Ir$_3$Ga$_9$. To investigate the magnetic contribution to the overall heat capacity, a non-magnetic, isostructural compound Y$_2$Ir$_3$Ga$_9$ was synthesized in single-crystal form following a similar procedure to that described above for Tb$_2$Ir$_3$Ga$_9$. The heat capacity of Y$_2$Ir$_3$Ga$_9$ does not show any magnetic ordering and hence can be taken as a basis for the lattice component of the heat capacity of Tb$_2$Ir$_3$Ga$_9$. The $T$-axis of the Y$_2$Ir$_3$Ga$_9$ specific heat data was scaled following a method developed by Bouvier et al~\citep{Bouvier.Cp.normalization} and later followed by others~\citep{Cp.Normalize2} (note: $\theta_D$ is 276 K from the single $\beta$ fit). Here the correction factor is calculated via Eq.~\ref{heat cap norm}

\footnotesize
\begin{equation}\label{heat cap norm}
\frac{\Theta_D (Tb_2Ir_3Ga_9)}{\Theta_D (Y_2Ir_3Ga_9)}=\left[\frac{2(M_Y)^\frac{2}{3}+3(M_{Ir})^\frac{2}{3}+9(M_{Ga})^\frac{2}{3}}{2(M_{Tb})^\frac{2}{3}+3(M_{Ir})^\frac{2}{3}+9(M_{Ga})^\frac{2}{3}}\right]^\frac{1 }{3},
\end{equation}
\normalsize 
where \textit{M$_x$} ($x$ = Y, Ga, Ir, Tb) is the atomic mass of each of the constituent atoms. In this way, a correction factor of 0.940 was calculated. The resulting data (Fig.~\ref{figure 4 resistivity}(b)) was then used to calculate the magnetic contribution to heat capacity, $C_{\rm mag}$, for Tb$_2$Ir$_3$Ga$_9$, where $C_{\rm mag} = C_{Tb}-C_{Y}$, which is shown in Fig.~\ref{figure 4 resistivity}c. Here, in addition to the $\lambda$-like transition, a broad feature is also evident. The magnetic contribution to the entropy is estimated by integrating, $\int{\frac{C_{\rm mag}}{T}}\, dT$. The resulting entropy is shown in Fig.~\ref{figure 4 resistivity}d. The maximum entropy we find is 17 J mol$^{-1}$Tb$^{-1}$K$^{-1}$, significantly smaller than the R\textit{ln}13 = 21.3 J mol$^{-1}$Tb$^{-1}$K$^{-1}$ expected for the $^7$F$_6$ ground state.

At the onset of antiferromagnetic order, we find $S = 5.75~{\rm Jmol^{-1}K^{-1}} = R \ln 2$ (Inset of Fig.~\ref{figure 4 resistivity}d). This can be understood via the splitting of the $^7$F$_6$ states of the Tb$^{3+}$ free ion by CEF into a ground state quasidoublet separated by a large ($>>$12 K) gap from the first excited state. While the CEF energy spectrum for \tig is not known, Tb compounds with similar site symmetry for the Tb$^{3+}$ ion, for example in TbAlO$_3$, place the ground state quasidoublet about 160 meV below the first excited state~\citep{CEF.TbAl03}. Similar behavior has been reported in TbNi$_2$Ge$_2$~\citep{Rln2}, where the Ising axis is the tetragonal $\vc$-axis. Here, a broad ``Schottky''-like feature was found in the magnetic heat capacity and attributed to thermal population of one or more CEF levels above the quasidoublet. A similar explanation likely applies to \tig. However, attempts to model the data above $T_{\rm N}$ to a Schottky form for a two-level system~\citep{schottky} lead to poor quality fits, probably reflecting the presence of groups of levels not adequately captured by a simple two-level expression. A better understanding of the CEF levels in \tig will be needed to model these data properly. The missing entropy in \tig, TbNi$_2$Ge$_2$ and other Tb$^{3+}$ containing systems~\citep{Cp.Normalize2,TbFe2.CEF,kumar2008magnetism} likely signals the existence of additional CEF levels at energies higher than that probed here.

\subsection*{Neutron Powder Diffraction}

\begin{figure}
\centering \includegraphics[width=1.0\linewidth]{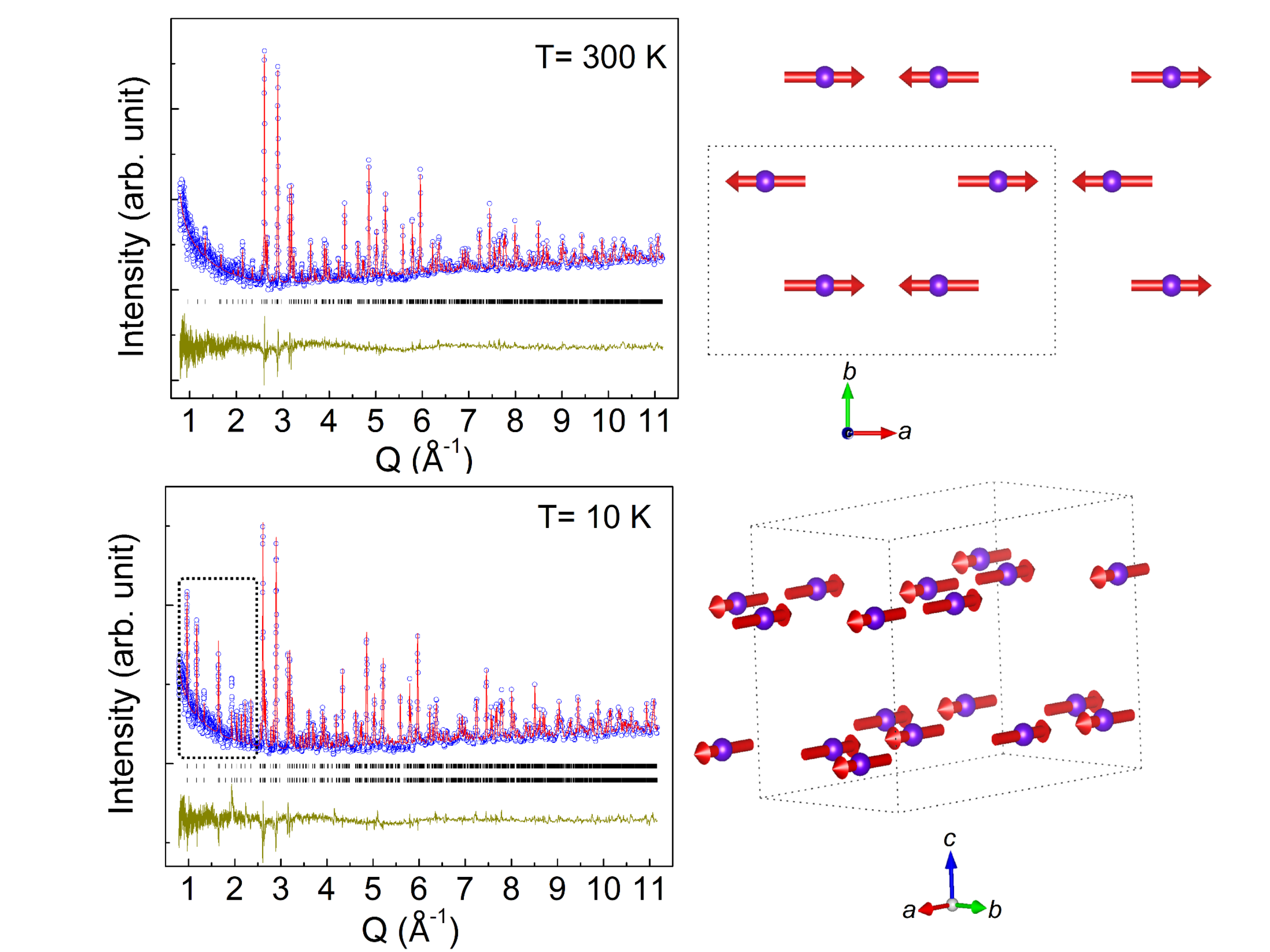} \caption{ 
Rietveld refinement of neutron powder diffraction patterns of Tb$_{2}$Ir$_{3}$Ga$_{9}$ at (a) 300 K, and (b) 10 K. The observed
data and the fit are indicated by the open circles and solid lines, respectively. The difference curve is shown at the bottom.
The vertical bars mark the positions of Bragg peaks for the nuclear phase (up) and magnetic phase 
(down). The dotted rectangle in (b) marks 7 strongest magnetic Bragg peaks. (c-d) Magnetic structure deduced from the refinement to the neutron data, showing the direction of the moments along $\va $-axis.}

\label{fig:Neutron} 
\end{figure}

The neutron powder diffraction pattern at 300 K, shown in Fig.~\ref{fig:Neutron}(a), evidences no secondary phases in the Tb$_{2}$Ir$_{3}$Ga$_{9}$ sample. Rietveld analysis confirms the orthorhombic structure with space group \textit{Cmcm} (No. 63) with the refinement goodness of $\chi^2\approx$3.5, as illustrated in the Fig.~\ref{fig:Neutron}(a). Upon cooling to 10 K ($<T_{\rm N}$), the intensities of more than 10 low-$Q$ peaks (Fig.~\ref{fig:Neutron}(b)) increase significantly, indicative 
of a magnetic contribution to the scattering. All magnetic reflections can be indexed on the 
nuclear (chemical) unit cell with a magnetic propagation vector
\textbf{k} = (0,0,0). The SARAH representational analysis program~\citep{Wills2000} 
was used to derive the symmetry-allowed magnetic structures. The symmetry allowed basis vectors for Tb sites are summarized in Table S1 of the SM~\citep{supp}. 
The neutron diffraction pattern is best fit using the $\Gamma$5, irreducible representations i.e., antiferromagnetic order with moment strictly along $\va $ axis with the refinement goodness of $\chi^2\approx$7.29.  Allowing a spin canting toward the $\vb $ axis does not improve the refinement. The ordered moment of Tb is found to be $7.5(2) \mu_{\rm B}$/Tb. Both results are 
consistent with the magnetization measurements. The magnetic structure
is displayed in Fig.~\ref{fig:Neutron}. It is worthwhile pointing out that Ir does not carry an ordered moment within the instrumental resolution. \\

\subsection*{XMCD}
\begin{figure}
\centering
\includegraphics[width=0.4\textwidth]{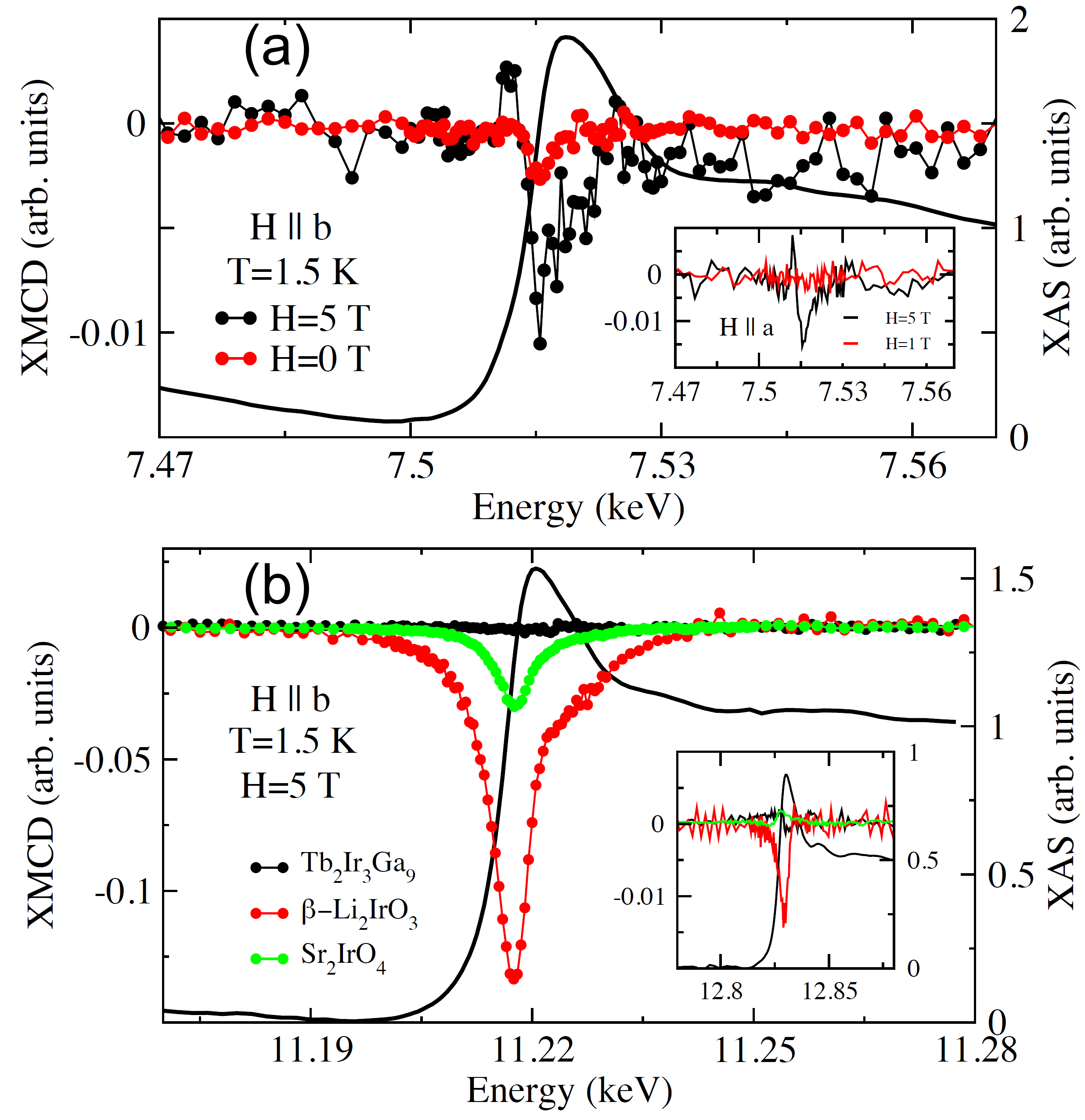}
\caption{XMCD and XAS on Tb and Ir L-edges at 1.5 K. (a) XMCD signals of Tb L$_3$-edge at 0 and 5 T field are plotted on the left axis while the XAS (solid black line) are plotted on the right axis for field parallel to crystallographic $\vb $-axis. Inset: XMCD signal at 1 and 5 T with field parallel to $\va $-axis. (b) XMCD signals of Ir L$_3$ edge at 5 T field are plotted on the left axis while the XAS (solid black line) are plotted on the right axis for field parallel to crystallographic $\vb $-axis. XMCD on Ir L$_3$ edge of $\beta$-Li$_2$IrO$_3$ and Sr$_2$IrO$_4$ are also plotted for comparison purposes. Inset: XMCD signal at 5 T of Ir L$_2$ edge.}
\label{fig: Tb edge}
\end{figure}


To more definitively explore the potential for Ir magnetism when the field is applied along the $\vb $-axis, we have performed 
X-ray magnetic circular dichroism (XMCD) measurements at beamline 4-ID-D of the Advanced Photon Source.  Data shown in the Fig.~\ref{fig: Tb edge} are averages of data sets collected with opposite applied field directions. While the Tb XMCD data at selected applied field values are consistent with the magnetometry data for both crystal orientations, no detectable XMCD signals were found at Ir L-edges. We can put an upper limit $<$ 0.01 $\mu_B$/Ir to the magnitude of any Ir magnetic moment by scaling to XMCD signals in $\beta$-Li$_2$IrO$_3$~\citep{XMCD1} (0.35 $\mu_B$/Ir) and Sr$_2$IrO$_4$~\citep{XMCD2} (0.05 $\mu_B$/Ir). This indicates that the finite ferromagnetic response along \textit{b}-axis is solely due to Tb moments either by a field induced canting of Tb spins or by a small but finite DM interaction between the nearest neighbors along the \textit{c}-direction, or a combination of the two. We will discuss these possibilities further in section~\ref{model}. The non-magnetic state of Ir is in accordance with our NPD data and the general consensus that the T atom in R$_2$T$_3$X$_9$ is magnetically inactive~\citep{DyCoAl}.

\subsection*{Electronic structure \& Magnetism}

Electronic structure calculations have been carried out within density functional theory (DFT) using the all-electron, full potential code WIEN2K~\citep{blaha2001wien} based on the augmented plane wave plus local orbital (APW + lo) basis set~\citep{sjostedt2000alternative}. The Perdew-Burke-Ernzerhof (PBE) version of the generalized gradient approximation (GGA)~\citep{PhysRevLett.77.3865} was chosen as the exchange correlation potential. Spin-orbit coupling (SOC) was introduced in a second variational procedure~\citep{singh2006planewaves}. The LDA+$U$ scheme improves over GGA or LDA in the study of systems containing correlated electrons by introducing the on-site Coulomb repulsion U applied to localized electrons (e.g., $4f$). We have performed calculations within the LDA+$U$ ensatz (using the fully localized version for the double-counting correction)~\citep{PhysRevB.49.14211} taking a reasonable $U$ value for this $f$-electron system (8 eV) comparable to the values obtained for TbN. A dense \textit{K}-mesh of 16$\times$16$\times$11, was used for the Brillouin zone sampling. An R$_{mt}$K$_{max}$ of 7 was chosen for all calculations. Muffin tin radii were 2.5 \textit{a.u.} for Ir and Tb, 2.7 \textit{a.u.} for Ga.

\begin{figure}
\centering
\includegraphics[width=0.5\textwidth]{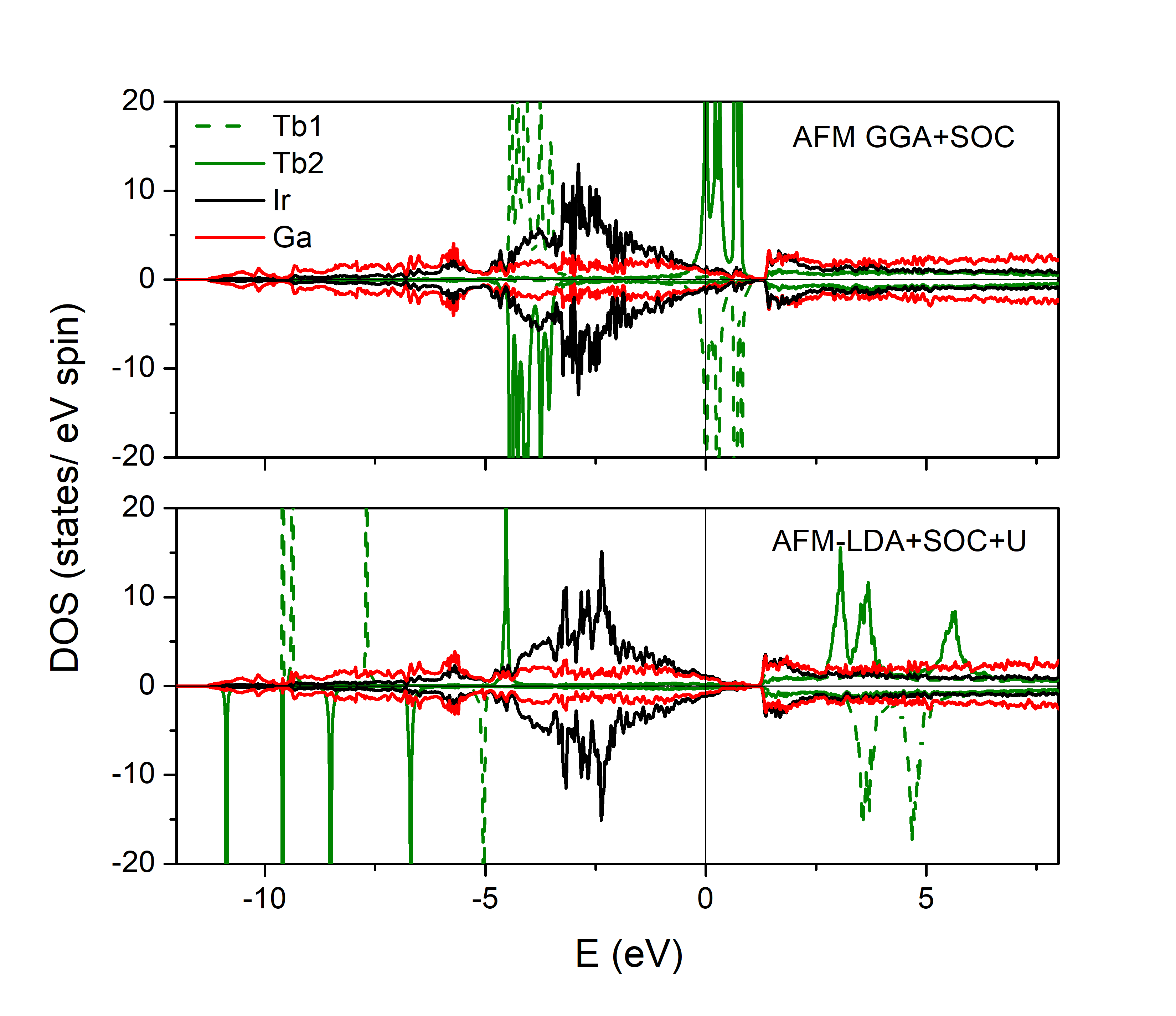}
\caption{Atom-resolved density of states for \tig. Left panel GGA, right panel LDA+U calculations.}
\label{fig: DOS}
\end{figure}

Using the experimental structure, DFT calculations were performed for a FM state, as well as for the collinear N\'eel AFM state proposed by neutrons (where each of the three types of bonds is AFM), and a collinear striped AFM phase. The AFM state proposed by neutrons is more stable than any other magnetic configuration tested by 16 meV/unit cell (u.c.). Once spin-orbit-coupling (SOC) is introduced, the preferred direction of the magnetization is the $\va$-axis with derived magnetocrystalline anisotropy energies MAE[100]-[010]= 0.35 meV/u.c., and MAE[100]-[001] = 0.75 meV/u.c.  The obtained $4f$ spin moment is  5.86 $\mu_{\rm B}$, and  the orbital moment 1.34 $\mu_{\rm B}$  (increased upon inclusion of a Coulomb $U$ to 6.07 and 1.40 $\mu_{\rm B}$, respectively). The total magnetic moment is then 7.2, and 7.5 $\mu_{\rm B}$  for GGA and LDA+$U$ calculations, respectively. Both agree with the ordered moment found by NPD and DC magnetization.

In the GGA calculation, as seen in Fig.~\ref{fig: DOS}, the minority spin channel for Tb atoms corresponding to their $f$ states, is pinned at the Fermi level (partially filled). Due to the highly localized character of the $4f$ electrons, it is unlikely that the density of states can have a finite Tb-$4f$ contribution at the Fermi level. In the LDA+U ground-state, there is no $4f$ weight at the Fermi level (all the weight is Ir-$d$ and Ga-$p$). There are instead different peaks of the $4f$-projected density of states well below and well above the Fermi level (shifted by the inclusion of a Coulomb $U$).

\section{Phenomenological Model of magnetism \lowercase{in} T\lowercase{b}$_2$I\lowercase{r}$_3$G\lowercase{a}$_9$}\label{model}

\begin{figure}
\begin{center}
\includegraphics[width=0.4\textwidth]{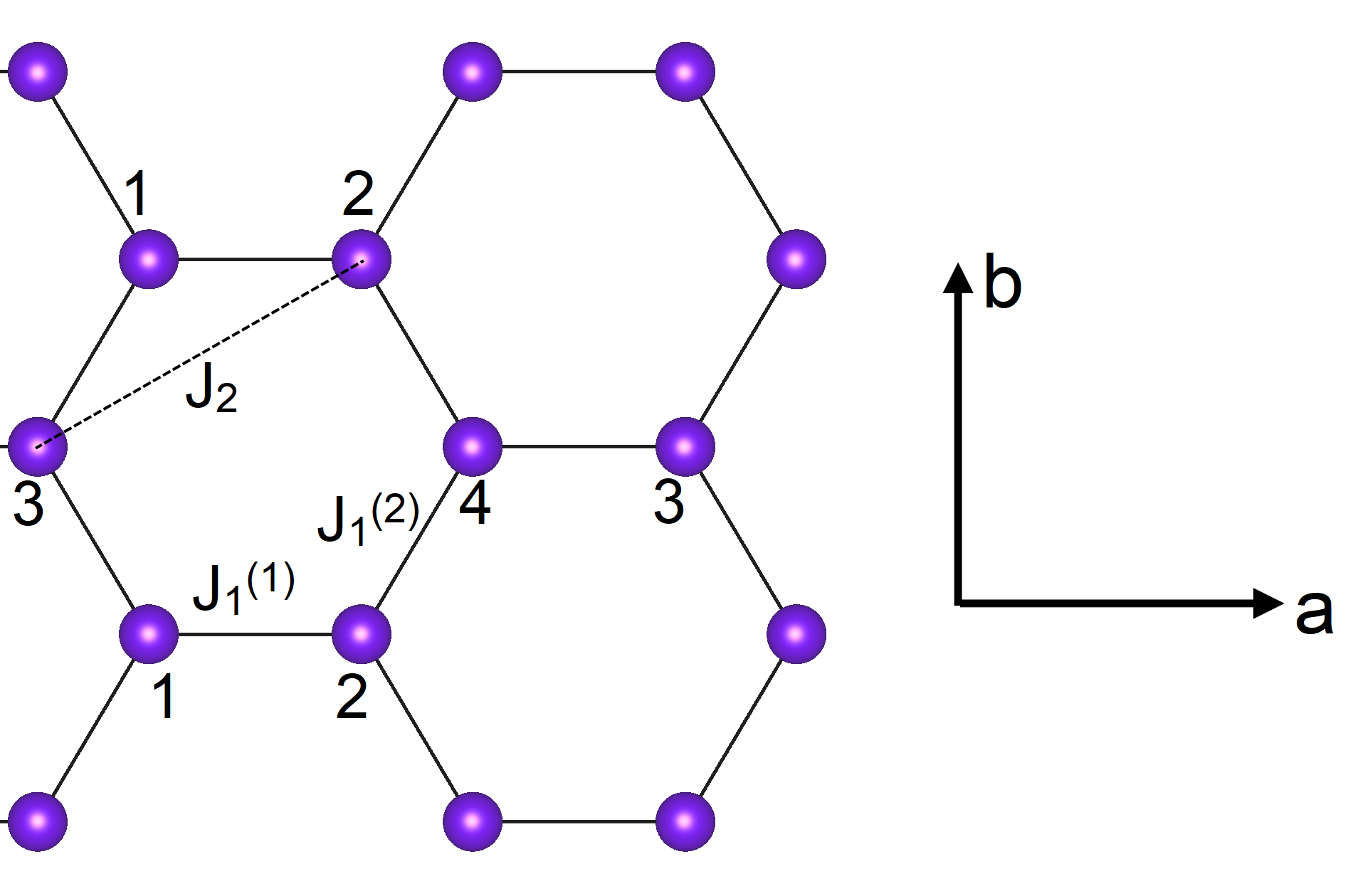}
\end{center}
\caption{Honeycomb layer of \tig with first-neighbor interactions $J_{1}^{(1)}$ and $J_{1}^{(2)}$ and second-neighbor interaction $J_2$.}\label{honey}
\end{figure}

A theoretical understanding of the magnetism in \tig confronts two constraints posed by the measurements discussed above. The first constraint posed is the absence of any phase transition when the field is applied along $\vb $ and $\vc$-axes, as confirmed by the measurements of the critical fields in applied fields of up to 60 T. Another constraint is provided by the observed scaling of the critical fields
when the external magnetic field is rotated by angle
$\theta $ away from the $\va $ axis in the $\va -\vb $ plane. As shown in Fig.~\ref{figure 3 MH}, measurements find that $B_{\rm c1}(\theta )\cos \theta $ 
and $B_{c2}(\theta )\cos \theta $ are roughly independent of $\theta $ up to $0.35 \pi $.  
This implies that only the component of the field
along $\va $ controls those phase transitions. Taken together, these two considerations suggest that the Tb moments can be approximately described as Ising spins aligned along $\pm \va $. However, a simple Ising model cannot faithfully capture all of the features discussed above, and we now build a phenomenological model in accordance with these considerations. We take the four Tb ions in the unit cell to have ``spins" 

\begin{equation}
\vS_i = S(\sin \theta_i \cos \phi_i ,\sin \theta_i \sin \phi_i, \cos \theta_i)
\end{equation}
with $S=6$, and ($\theta$,$\phi$)$_i$ are spherical polar coordinates.  Elastic neutron measurements indicate that the zero-field state of the Tb ions has $\theta_i = \pi/2$ with
$\phi_1 \approx \phi_4 \approx 0$ and $\phi_2 \approx \phi_3 \approx \pi $. 

\begin{figure}
\begin{center}
\includegraphics[width=0.5\textwidth]{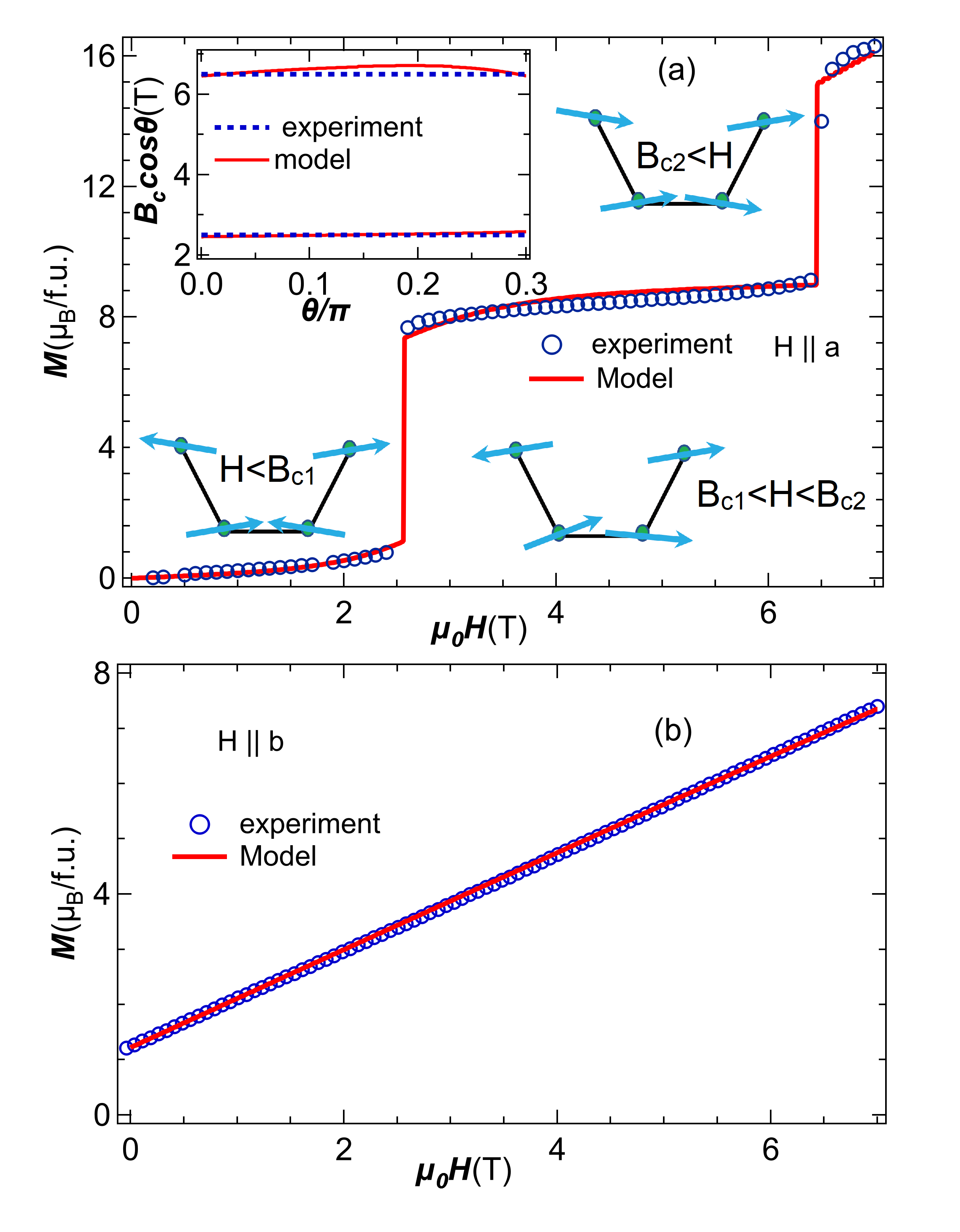}
\end{center}
\caption{Predicted and measured magnetization with field along (a) $\va $, and (b) $\vb $ axes. Inset: the predicted variation of $B_{\rm c} \cos \theta $ where $\theta $ is the angle of the field in the $xy$ plane. Dashed lines are the experimental data.}
\label{unitcell}
\end{figure}
The Hamiltonian is given by
\begin{eqnarray}
{\cal H}&=&-\frac{1}{2}\sum_{i, j}^{1st} J_{1, ij}\, \vS_i \cdot \vS_j  -\frac{1}{2}\sum_{i, j}^{1st} J_{1, ij}^n \, \vS_i \cdot \vn_{ij}  \, \vS_j \cdot \vn_{ij}   \nonumber \\
&-& \frac{1}{2}J_2 \sum_{i,j}^{2nd} \vS_i \cdot \vS_j
- K_2 \sum_i {S_{ia}}^2  -K_p \sum_i {S_{ic}}^2 
\nonumber \\ &-&
\frac{1}{2} K_h \sum_i \Bigr\{ (S_{ia} +iS_{ib})^6 +(S_{ia}-iS_{ib})^6\Bigl\} \nonumber \\
&-&\frac{1}{2}\sum_{i, j}^{1st} {\bf D}_{ij} \cdot (\vS_i \times \vS_j )
- g\mb \vB \cdot \sum_i \vS_i,
\label{ham}
\end{eqnarray}
where the field is along the $\alpha =a$, $b$, or $c$ direction and 
the exchange interactions are indicated in Fig.~\ref{honey}.  The factors of $1/2$ avoid double counting.

First neighbors are coupled by both isotropic $J_{1, ij}$ and directional $J_{1, ij}^n$ exchange couplings.
The latter couples spins along the direction of the bond so that 
\begin{equation}
\vn_{ij} = \frac{\vR_j - \vR_i}{\vert \vR_j - \vR_i\vert }
\end{equation}
is a unit vector from site $i$ to site $j$.  
To account for the orthorhombic distortion of the honeycomb lattice, we further break 
$J_{1, ij}$ and $J_{1, ij}^n$ into two parts:  $J_1^{(1)}$ and $J_1^{n(1)}$ acts between sites 1 and 2 or 3 and 4 along the $a$ axis
while $J_1^{(2)}$ and $J_1^{n(2)}$ act between sites 1 and 3 or 2 and 4 at 60$^\circ $ from the $a$ axis.  
Second neighbors 1 and 4 or 2 and 3 are coupled by $J_2$.
The easy-axis anisotropy $K_2$ aligns the spins along the $a$ axis due to the orthorhombic distortion of the lattice and the easy-plane anisotropy $K_p < 0$ keeps the spins in the $a-b$ plane.  The hexagonal anisotropy $K_h > 0$ favors the spins $\vS_i $ to lie along the three pseudo-hexagonal axis at $\phi_i =0$ and $\pm \pi/3$.  In terms of the spin angles, this energy can be writtten $-S^6 K_h \cos 6\phi_i $ at each site.

The DM interactions ${\bf D}_{ij}$ along $\vc $ act only between first neighbors (with opposite signs between sites 1 and 2 and between 3 and 1 or 2 and 4) and produces the zero-field canted moment along $\vb $. This DM interaction is allowed by the alternating positions of the Ir ions around each Tb-Tb bond.

This complex model is required to ``tame" the magnetization so that no phase transition occurs when the field is applied along $\vb $.
A simpler model that neglects the directional exchange has a $\chi^2$ about 16 times greater.
The best fits for this model are shown in Fig.~\ref{unitcell}. Notice that this model describes the experimental measurements in all three phases at 1.8 K. The $\vc$-axis behavior, not shown, is a good fit to the data shown in Fig.~\ref{figure 3 MH}b. The fits for the model were constrained by the requirement that 
$B_{cn}(\theta )\cos \theta $ ($n=1$ or 2) are approximately constant as a function of $\theta $, as measured experimentally. The scaled critical fields from the model are plotted as a function of $\theta $ in the inset of Fig.~\ref{unitcell}. The scaled fields deviate from their $\theta =0$ values only above about $0.3\pi $. We have neglected the presumably weak coupling between planes.  While it is not known how the neighboring planes are magnetically configured, if they respond identically to the applied fields, then their exchange coupling will not change with field.


The fitted values of the parameters for the model are given in Table~\ref{obmodes}. The largest parameter is the easy-plane anisotropy $K_p \approx -0.88$ meV, which keeps the Tb spins in the $a-b$ plane. The weak hexagonal anisotropy $K_h$ favors the spins to lie along the three hexagonal axis. 
Notice that all the exchange parameters are AFM. While the first-neighbor exchange between spins 1 and 2 or 3 and 4 contains both isotropic $J_1^{(1)}$ and 
directional $J_1^{n(1)}$ contributions, the exchange between spins 1 and 3 or 2 and 4 is primarily directional with $\vert J_1^{(2)}\vert \ll \vert J_1^{n (2)}\vert $. The next nearest neighbor interaction $J_2 <$ 0 stabilizes the intermediate metamagnetic phase. With $J_2$=0, B$_{\rm c2}$ = B$_{\rm c1}$ and the intermediate phase would be absent. We note that the extracted parameters are consistent with the observed transition temperature with $J \approx$ 0.025 meV and the mean-field transition temperature ($z$/3)$J$ $S$($S$+1) = 1 meV or 11.6 K.

\begin{table}
\caption{Parameters in meV}
\begin{tabular}{|cc|cc|cc|cc|}
\hline
 parameter && value &\\
\hline
$J_1^{(1)}$ && -0.016 &\\
$J_1^{(2)}$ &&  -0.0004 &\\
$J_1^{n (1)}$ && -0.024 &\\
$J_1^{n (2)}$ &&  -0.063 &\\
$J_2$ &&  -0.0081 &\\
\hline
$K_2$ && 0.050 &\\
$K_p$ &&  -0.88 &\\
$K_h$ &&  $1.4 \times 10^{-7}$ &\\
\hline
$D$ &&  0.0065 &\\
\hline
\end{tabular}
\label{obmodes}
\end{table}

\section{Conclusion}
We have synthesized single crystals of the honeycomb lattice antiferromagnet \tig. The observed magnetism is highly anisotropic with an AFM transition at T$_N\approx$12.5 K. Two step-like metamagnetic transitions were found when the magnetic field was applied along the magnetic easy $\va$-axis, reflecting the Ising nature of the Tb$^{3+}$ quasidoublet. Neutron powder diffraction revealed the direction of the magnetic moment along the $\va$-axis, in accordance with the magnetization data. A broad peak found for the $\vc$-axis susceptibility is attributed due to CEF effects, as is a similar broad maximum in the magnetic specific heat above the N\'eel transition. A phenomenological model was proposed that describes all of the magnetic data well, including the angle-dependence of the metamagnetic transitions. A small but finite DM interaction between nearest neighbors in Tb-Tb planes, which acts along $\vc$-axis, was found to be essential in describing the observed scaling behavior of the metamagnetic transitions.  Notably, the inclusion of a bond directional anisotropy to the magnetic exchange is essential to proper modeling of the data, highlighting that \tig joins the family of honeycomb magnets with such anisotropic exchange. Beyond this, the Ising behavior of the Tb moments and the honeycomb lattice arrangements of the Tb atoms make this compound a fertile ground to investigate the interplay among various magnetic interactions and crystal field effects. As such, field dependent neutron scattering experiments on single crystals are a logical next step toward understanding this honeycomb lattice system.

\acknowledgments 
This work was sponsored by the U.S. Department of Energy, Office of Science, Basic Energy Sciences, Materials Sciences and Engineering Division. A portion of this research used resources at Spallation Neutron Source, a DOE Office of Science User Facility operated by the Oak Ridge National Laboratory. Work performed at the National High Magnetic Field Laboratory, USA, was supported by NSF Cooperative Agreements DMR-1157490 and DMR-1644779, the State of Florida, U.S. DoE, and through the DoE Basic Energy Science Field Work Project Science in 100 T. Work at the APS was supported by the U.S. Department of Energy, Office of Science, under Contract No. DE-AC02-06CH11357. A.S.B. thanks ASU for startup funds. The authors would like to thank Dr. David Parker, ORNL, for useful discussions.

\bibliography{TbBib}

\end{document}